\documentclass[conference,11pt]{IEEEtran}
\usepackage{cite}
\usepackage{amsmath,amssymb,amsfonts}
\usepackage{algorithm}
\usepackage{algpseudocode}
\usepackage{float}
\usepackage{graphicx}
\usepackage{array}
\usepackage{tabularx}
\usepackage{textcomp}
\usepackage{xcolor}
\usepackage[hyphens]{url}
\usepackage{microtype}
\usepackage{graphicx}
\usepackage{multirow}
\usepackage{pgfplots}
\usepackage{comment}
\usepackage{makecell}
\usepackage{stfloats}
\usepackage{tikz}
\usepackage{float}
\usepackage{amsmath}
\usepackage{subfigure}
\usepackage{booktabs}

\def\BibTeX{{\rm B\kern-.05em{\sc i\kern-.025em b}\kern-.08em
    T\kern-.1667em\lower.7ex\hbox{E}\kern-.125emX}}

\title{ReLMXEL: Adaptive RL-Based Memory Controller with Explainable Energy and Latency Optimization}

\author{
    \parbox{\textwidth}{
        \centering

        \parbox[t]{0.45\textwidth}{
        \centering
        \textbf{Panuganti Chirag Sai}\\
        \textit{\normalsize Department of Mathematics and Computer Science}\\
        \textit{\normalsize Sri Sathya Sai Institute of Higher Learning}\\
        chiragsaipanuganti@sssihl.edu.in
        }
        \hfill
        \parbox[t]{0.45\textwidth}{
        \centering
        \textbf{Gandholi Sarat}\\
        \textit{\normalsize Department of Mathematics and Computer Science}\\
        \textit{\normalsize Sri Sathya Sai Institute of Higher Learning}\\
        gandholisarat@sssihl.edu.in
        }

        \vspace{0.5cm}

        \parbox[t]{0.45\textwidth}{
        \centering
        \textbf{R. Raghunatha Sarma}\\
        \textit{\normalsize Department of Mathematics and Computer Science}\\
        \textit{\normalsize Sri Sathya Sai Institute of Higher Learning}\\
        rraghunathasarma@sssihl.edu.in
        }
        \hfill
        \parbox[t]{0.45\textwidth}{
        \centering
        \textbf{Venkata Kalyan Tavva}\\
        \textit{\normalsize Department of Computer Science and Engineering}\\
        \textit{\normalsize Indian Institute of Technology Ropar}\\
        kalyantv@iitrpr.ac.in
        }

        \vspace{0.2cm}

        \parbox[t]{0.5\textwidth}{
        \centering
        \textbf{Naveen M}\\
        \textit{\normalsize AI Performance Engineer}\\
        \textit{\normalsize Red Hat}\\
        nmiriyal@redhat.com
        }

    }
}

\begin{document}
\maketitle
\thispagestyle{plain}
\pagestyle{plain}

\begin{abstract}
    Reducing latency and energy consumption is critical to improving the efficiency of memory systems in modern computing. This work introduces ReLMXEL (Reinforcement Learning for Memory Controller with Explainable Energy and Latency Optimization), a explainable multi-agent online reinforcement learning framework that dynamically optimizes memory controller parameters using reward decomposition. ReLMXEL operates within the memory controller, leveraging detailed memory behavior metrics to guide decision-making. Experimental evaluations across diverse workloads demonstrate consistent performance gains over baseline configurations, with refinements driven by workload-specific memory access behaviour. By incorporating explainability into the learning process, ReLMXEL not only enhances performance but also increases the transparency of control decisions, paving the way for more accountable and adaptive memory system designs.
\end{abstract}

\section{Introduction}
    In modern computing systems, Dynamic Random Access Memory (DRAM) is de-facto memory technology and plays a critical role in overall system performance, especially for memory- and compute-intensive workloads such as those encountered in machine learning (ML) training and inference. Consequently, significant research focuses on improving DRAM efficiency, particularly in reducing latency and energy consumption. The memory controller managing the communication between the processor and DRAM, is pivotal in achieving these optimizations. A survey by Wu et al.~\cite{survey} reviews the growing use of machine learning in computer architecture, highlighting reinforcement learning (RL) as a promising technique for designing self optimizing memory controllers. These controllers are modeled as RL agents that choose DRAM commands based on long-term expected benefits and incorporate techniques such as genetic algorithms and multi-factor state representations to handle diverse objectives like energy and throughput. One prominent approach is the self-optimizing memory controller proposed by Ipek et al.~\cite{Selfoptimising}, that uses RL to adapt scheduling decisions and outperform static policies across various workloads.
    
    Despite these improvements, there is a lack of transparency in RL-driven decisions, hindering their adoption in real-world systems that require explainability, reliability and trust. To bridge this gap, we introduce Reinforcement Learning for Memory Controller with Explainable Energy and Latency Optimization (ReLMXEL), a novel multi-agent RL-based memory controller. ReLMXEL dynamically tunes memory policies to optimize latency and energy across diverse workloads, including several that exhibit computational patterns commonly found in machine learning (ML) applications, such as dense linear algebra (GEMM), memory-bound operations (STREAM, mcf) and irregular data access patterns (BFS, omnetpp) while incorporating explainability techniques to make its decisions interpretable. This approach builds upon prior work in adaptive memory systems and aims to balance performance with accountability in complex computing environments.

\section{Literature Review}
        \vspace*{-0.3cm}
        \begin{figure}[H]
            \begin{center}
            \includegraphics[width=0.6\textwidth, trim=4cm 6cm 0cm 3.6cm, clip]{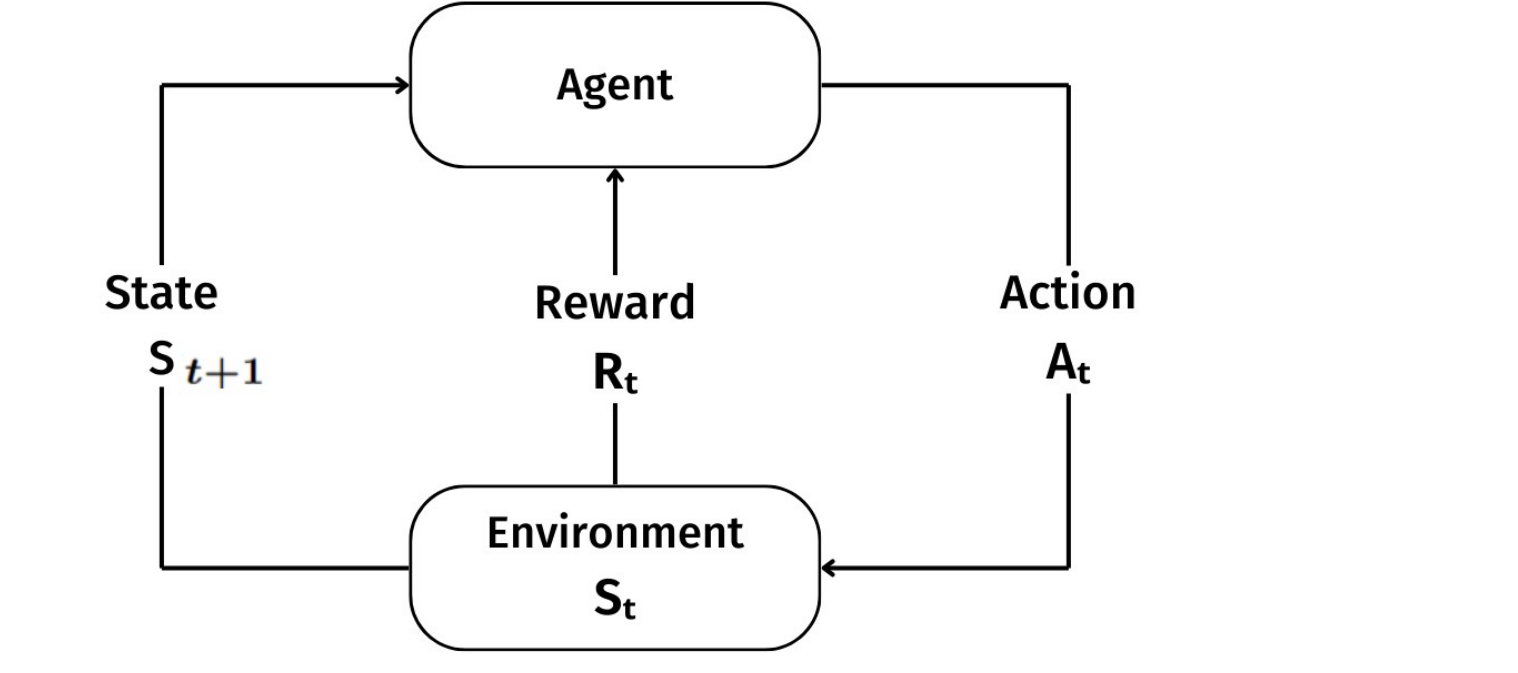}
                \caption{Reinforcement Learning Framework~\cite{rlanintroduction}}
                \vspace*{-0.5cm}
                \label{fig:General}
            \end{center}
        \end{figure}
        In a RL framework, an agent interacts with the environment over discrete timesteps. At each timestep $t$, the agent observes the current state $S_t$, selects an action $A_t$ based on a policy $\pi(a|s)$, receives a reward $R_t$, and transitions to a new state $S_{t+1}$. The goal of the agent is to learn an optimal policy that maximizes the expected cumulative reward over time. This process continues iteratively, allowing the agent to learn a policy $\pi(a|s)$ that maximizes the expected cumulative reward over time.

        Machine learning approaches generally require large, labeled datasets and assume that data distributions remain stationary. However, memory systems exhibit highly dynamic behavior, with workloads and access patterns changing rapidly over time. Traditional ML methods lack the capability to adapt on-the-fly and cannot effectively capture the dynamism in memory systems. Whereas, an RL agent learns through direct interaction with the environment, making decisions based on real-time feedback rather than relying on pre-collected data. This allows RL to effectively handle non-stationary environments by continuously adapting its policy as system conditions evolve. Additionally, RL optimizes long-term cumulative rewards, and supports multi-objective optimization tasks such as balancing energy efficiency, bandwidth, and latency. These strengths make RL particularly well-suited for memory controller parameter tuning.

    \subsection{Self-Optimizing Memory Controllers: A Reinforcement Learning Approach}
        The Self-Optimizing Memory Controller by Ipek et al.\cite{Selfoptimising} overcomes the limitations of static DRAM controllers by using reinforcement learning to dynamically adapt command scheduling. It models the controller as an RL agent interacting with an environment composed of processor cores, caches, buses, DRAM banks, and scheduling queues. The state includes features such as read/write counts and load misses, while actions include Precharge, Activate, Read-CAS, Write-CAS, REF, or NOP commands. The agent receives a reward of 1 for read/write and 0 otherwise. SARSA\cite{rummerysarsa,rlanintroduction} updates Q-values~\cite{Q-Table} using a Cerebellar Model Articulation Controller (CMAC) function approximator~\cite{cmac} with overlapping coarse-grained Q-tables to handle the large state space. This approach enables adaptability to workload changes, optimizing scheduling decisions. However, it focuses solely on scheduling, neglecting important parameters such as arbitration, refresh policies, page policies, scheduler buffer policies, and the maximum number of permitted active transactions. Furthermore, the lack of explainability in learned policies limits interpretability and reliability, highlighting the need for memory controllers that balance adaptability with transparency.

    \subsection{Pythia: A Customizable Hardware Prefetching Framework Using Online Reinforcement Learning}
        The Pythia~\cite{Bera_2021} framework proposes a prefetcher for cache optimization using reinforcement learning. Pythia treats the prefetcher as an RL agent, where, for each demand request, it observes various types of program context information to make a prefetch decision. After each decision, Pythia receives a numerical reward that evaluates the quality of the prefetch, considering current memory bandwidth usage. This reward strengthens the correlation between the observed program context and the prefetch decisions, helping generate more accurate, timely, and system-aware prefetch requests in the future. The primary objective of Pythia is to discover the optimal prefetching policy that maximizes the number of accurate and timely prefetch requests while incorporating system-level feedback. The state space is a \( k \)-dimensional vector of program features, \( S \equiv \{\varphi_1^S, \varphi_2^S, \dots, \varphi_k^S\} \). The action is the selection of a prefetch offset from a set of pre-determined offsets. The reward is calculated based on factors like Accurate and Timely, Accurate but Late, Loss of Coverage, Inaccurate, and No Prefetch~\cite{Bera_2021}.

    \subsection{Reinforcement Learning using Reward Decomposition}
    \label{RL_RD}
        In \textit{Explainable Reinforcement Learning via Reward Decomposition}~\cite{juozapaitis2019rewarddecomposition}, the scalar reward in conventional reinforcement learning is decomposed into a reward vector, where each element represents the reward from a specific component. Say, we have two possible actions \( a_1 \) and \( a_2 \) available to the agent in a given state \( s \). The reward vector helps explain why an action \( a_1 \) is preferred over another \( a_2 \) in a state \( s \). The explanation is provided through the \textit{Reward Difference Explanation (RDX)}, defined as:
        \begin{equation}
            \Delta(s, a_1, a_2) = \vec{Q}(s, a_1) - \vec{Q}(s, a_2),
            \label{eq:rdx}
        \end{equation}

        wherein, each component \( \Delta_c(s, a_1, a_2) \) represents the difference in expected return with respect to a component \( c \). A positive \( \Delta_c \) indicates an advantage of \( a_1 \) over \( a_2 \), and vice versa. When the reward components are numerous, the authors introduce \textit{Minimal Sufficient Explanation (MSX)}. An MSX is a minimal subset of components whose cumulative advantage justifies the preference of one action over another. Specifically, an MSX for \( a_1 \) over \( a_2 \) is given by the smallest subset \( \text{MSX}^+ \subseteq \mathcal{C} \) such that:
        \begin{equation}
            \sum_{c \in \text{MSX}^+} \Delta_c(s, a_1, a_2) > d,
            \label{eq:msxplus}
        \end{equation}
        where \( d \) is the total disadvantage from negatively contributing components:
        \begin{equation}
            d = -\sum_{\Delta_c(s, a_1, a_2) < 0} \Delta_c(s, a_1, a_2).
            \label{eq:disadvantage}
        \end{equation}

        To verify whether each component in \( \text{MSX}^+ \) is necessary, a necessity check is introduced and calculated as:
        \begin{equation}
            v = \sum_{c \in \text{MSX}^+} \Delta_c(s, a_1, a_2) - \min_{c \in \text{MSX}^+} \Delta_c(s, a_1, a_2),
            \label{eq:justinsufficient}
        \end{equation}

        Finally, checking if any subset of negative components \( \text{MSX}^- \) has a total disadvantage exceeding \( v \), if so, all the elements in \( \text{MSX}^+ \) are deemed necessary, leading to the formal definition:
        \begin{equation}
            \text{MSX}^- = \arg\min_{M \subseteq \mathcal{C}} |M| \text{ s.t. } \sum_{c \in M} -\Delta_c(s, a_1, a_2) > v
            \label{eq:msxminus}
        \end{equation}

\section{ReLMXEL}
        \begin{figure}[ht]
            \hspace{-1.7cm}
            \includegraphics[width=0.7\textwidth, trim=-0.5cm 6cm 0cm 6.8cm, clip]{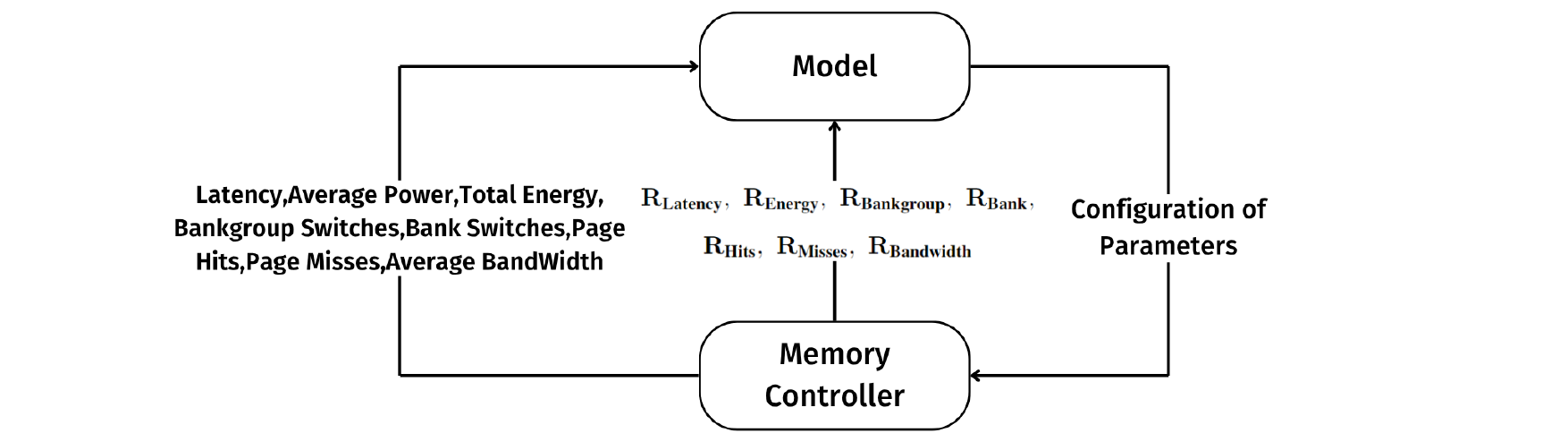}
            \caption{ReLMXEL Framework}
            \vspace*{-0.5cm}
            \label{fig:ReLMXEL}
        \end{figure}
        
        We now propose a strategy: Reinforcement Learning for Memory Controller with Explainable Energy and Latency Optimization (ReLMXEL), that operates within an RL setting. The memory controller serves as the environment, providing information/metrics such as latency, average power, total energy consumption, bandwidth utilization, bank and bankgroup switches, and row buffer (page) hits and misses. Latency is tracked per request to reflect internal delays. Average power and total energy are derived from DRAM state transitions and activity counters. The bandwidth utilization captures interface efficiency. The bank and bank group switches are logged to monitor access locality, and the row buffer hits and misses indicate the effectiveness of row management. These metrics provide deep visibility into DRAM behavior and serve as observations for the RL agent, which computes per-metric rewards and selects actions to optimize the overall DRAM performance.

        \begin{algorithm}[t]
            \caption{ReLMXEL Algorithm}
            \label{alg:relmxel}
            \begin{algorithmic}[1]
                \State \textbf{Input:} Timesteps $T$, base seed $s$, threshold $w$, learning rate $\alpha$, discount factor $\gamma$
                \State \textbf{Output:} All Q-tables $Q_i$ and $\mathcal{R}_{\text{C}}$
                \State Initialize $\epsilon_{\text{old}}$, $\epsilon_{\text{new}}$, $\mathcal{R}_{\text{C}} \gets 0$
                \For{$i = 1$ to $N$} \Comment{\textbf{$N$ agents}}
                    \State $s_i \gets s + i$ \Comment{\textbf{Seed per agent}}
                    \State Initialize $Q_i(s, a_i, r)$
                \EndFor
                \State Initialize current state $s_{\text{old}}$
                \State Select initial action vector $\mathbf{a} \gets (a_1, \dots, a_N)$ using \\ $\epsilon$-greedy strategy
                \For{$t = 1$ to $T$}
                    \State Apply action $\mathbf{a}$ to memory controller
                    \State Extract performance metrics $\left( R_{j,\text{obs}} \right)_{j=1}^M$
                    \vspace*{0.02cm}
                    \State Compute rewards metric-wise using Eq.~\eqref{eq:reward}
                    \If{$t < w$}
                        \State $\epsilon \gets \epsilon_{\text{old}}$
                    \Else
                        \State $\epsilon \gets \epsilon_{\text{new}}$
                        \State $\mathcal{R}_{\text{C}}\gets \mathcal{R}_{\text{C}}+R_{\text{T}}$ \hfill \footnotesize \Comment{\textbf{Cumulative Reward}}
                        \normalsize
                    \EndIf
                    \For{$i = 1$ to $N$} \Comment{Each agent chooses action}
                        \If{random number $< \epsilon$}
                            \State $a'_i \gets$ random action for agent $i$
                        \Else
                            \State $a'_i \gets \arg\max_{a'_i} \sum_j Q_i(s_{\text{old},i}, a'_i, r_j)$
                        \EndIf
                    \EndFor
                    \State $\mathbf{a}'\gets(a'_1, a'_2, \dots, a'_N)$ \Comment{\textbf{Next Action}}
                    \State $s_{\text{new}} \gets \mathbf{a}$ \hfill \Comment{\textbf{New State}}
                    \For{$i = 1$ to $N$}
                        \For{each reward $r_j$}
                            \State Compute $Q_i(s_{\text{old},i}, a_i, r_j)$ using Eq.~\eqref{eq:sarsa_update}
                        \EndFor
                    \EndFor
                    \State $s_{\text{old}} \gets s_{\text{new}}$
                    \State $\mathbf{a} \gets \mathbf{a}'$
                \EndFor
                \State \Return All Q-tables $Q_i$, $\mathcal{R}_{\text{C}}$
            \end{algorithmic}
        \end{algorithm}

        The actions consist of configurable DRAM parameters, including \texttt{PagePolicy} (Open, OpenAdaptive, Closed, ClosedAdaptive), which governs whether a row remains open or closed immediately after access. \texttt{Scheduler} (FIFO, FR-FCFS, FR-FCFS Grp), defines how memory requests are prioritized and ordered to balance fairness and throughput. \texttt{SchedulerBuffer} (Bankwise, ReadWrite, Shared), determines how request queues are organized, either by bank, by read/write separation, or as a shared buffer. \texttt{Arbiter} (Simple, FIFO, Reorder), selects which commands proceed to DRAM based on fixed priorities, order, or dynamic reordering to improve timing efficiency. \texttt{RespQueue} (FIFO, Reorder), controls the order in which responses are sent back to the requester. \texttt{RefreshPolicy} (NoRefresh, AllBank), manages how DRAM refresh operations are performed to maintain data integrity while minimizing interference. \texttt{RefreshMaxPostponed} ($0,\dots,7$), and \texttt{RefreshMaxPulledin} ($0,\dots,7$), allow the controller to delay or advance refreshes within limits to reduce conflicts with memory accesses. \texttt{RequestBufferSize} limits the number of outstanding requests the controller can hold and \texttt{MaxActiveTransactions} ($2^x$ where $x=0,\dots,7$) controls the number of concurrent active DRAM commands. Through iterative interaction, the agent learns to tune DRAM parameters for optimal efficiency. It can be noted that the framework is generalized and can be extended/adapted to various standards (DDR/GDDR/LPDDR, etc.,) and generations, and varying polices like SameBank Refresh,chopped-BurstLength, etc.

        As described in Algorithm~\ref{alg:relmxel}, each configurable parameter is associated with a Q-table~\cite{Q-Table}. The reward is calculated by the function:
        \begin{equation}
            R_{X} = \frac{R_{\mathrm{target}}}{|R_{\mathrm{target}} - R_{\mathrm{observed}}|}
            \label{eq:reward}
        \end{equation}
        wherein, the subscript {\it X} corresponds to the reward \textit{R} of a performance metric, $R_{\text{target}}$  and $R_{\text{observed}}$ corresponds to the ideal reward and the reward of current timestep respectively. $R_{\text{T}}$ is defined as:
        \begin{equation}
            R_T = \sum_{i=1}^{7} R_{X_i} 
            \ \text{; where } X_i \text{ is a performance metric}
            \label{eq:total_reward}
        \end{equation}
        The Q-value~\cite{Q-Table}, denoted as \( Q(s, a) \), represents the expected cumulative reward for taking an action \( a \) in the state \( s \) and following the current policy. These Q-values~\cite{Q-Table} are stored in a Q-table, a lookup table organized such that each dimension corresponds to discrete states and possible actions for specific DRAM parameters. During decision-making, the agent uses the current state and possible actions as indices to retrieve the associated Q-values, enabling efficient evaluation of expected rewards.
        The model follows the \textit{SARSA}~\cite{rlanintroduction,rummerysarsa} update rule to continuously improve its policy based on observed transitions.
        \begin{equation}
            Q(s_{t}, a_{t})\gets Q(s_{t}, a_{t})+\alpha\Big[r_{t}+\gamma Q(s_{t+1},a_{t+1})-Q(s_{t}, a_{t})\Big]
            \label{eq:sarsa_update}
        \end{equation}
        where \( s_t \) and \( a_t \) are the current state and action, \( r_t \) is the received reward, \( s_{t+1} \) is the next state, and \( a_{t+1} \) is the next action chosen using the current policy. Here, \( \alpha \) is the learning rate (\( 0 < \alpha \leq 1 \)) and \( \gamma \) is the discount factor (\( 0 \leq \gamma \leq 1 \)).
        
        To guide the learning process, we define a warmup threshold \( w \), representing the initial number of iterations focused on exploration, this allows the algorithm to adequately explore various memory controller parameters before commencing the optimization. A base seed is used to generate a unique seed for each agent.
        
    \subsection{Explainability of ReLMXEL}
        Following the approach given by Juozapaitis et al., in ReLMXEL, the conventional scalar RL reward is decomposed into a vector representing system-level performance metrics. The Q-function is decomposed into individual Q-values for each of the reward types. For a given state \( s \), an action \( a_1 \) is selected over \( a_2 \) iff:
        \begin{equation}
        \sum_c Q_c(s, a_1) > \sum_c Q_c(s, a_2)
        \label{eq:act_selection}
        \end{equation}
        To understand further we use RDX. But this setup leads us to consider every possible action state pair. To simplify, we apply the MSX as in~\ref{RL_RD} which provides a rationale for selecting action \( a_1 \) over \( a_2 \) if
        \begin{equation}
            \sum_{c \in \text{MSX}^+} \Delta_c(s, a_1, a_2) > d
            \label{eq:msxplus}
        \end{equation}
        where d is the disadvantage from negatively contributing factors.
        
        Consider an action \( a_1 \) which uses open page policy and improves latency and bandwidth but negatively impacts energy. Another action \( a_2 \) which uses closed page policy offers huge improvement in energy but negatively impacts latency and bandwdith. MSX identifies the smallest subset of components that adequately justifies the preference for \( a_2 \). For example, if the energy improvement is substantial enough to outweigh the latency and bandwidth drawbacks, MSX helps explain the decision as 'the improvement in energy alone justifies the action, despite losses in other components'.

        Similarly, consider an action \( a_3 \) that uses simple arbitration policy and reduces energy consumption significantly but negatively impacts the latency and bandwidth usage. On the other hand, another action \( a_4 \) using reorder arbitration policy provides moderate improvements in both latency and bandwidth with slight increase in energy consumption. MSX could justify the action \(a_3\) by explaining: 'The significant reduction in energy consumption is enough to justify \(a_3\) against moderate improvements in latency and bandwidth of \(a_4\).'

\section{Experimental Setup and Results}
    We performed experiments using DDR4 memory~\cite{jedecDDR4} in DRAMSys simulator~\cite{DRAMSys}, featuring a burst length of eight, four bank groups with four banks each, and each bank comprising of 32,768 rows and 1024 columns of size 8 bytes per device. The system uses a single channel, single rank configuration, made up of $x8$ DRAM devices. The baseline memory controller employs an OpenAdaptive Page Policy, outperforming static open and closed policies~\cite{intel2024openpage}, and uses the widely adopted FR-FCFS scheduling~\cite{FRFCFS} algorithm with a bank wise scheduler buffer supporting up to eight requests. It also supports an All-bank refresh policy with up to eight postponed and eight pulled-in refreshes. The controller manages up to 128 active transactions, and an arbitration unit reorders incoming requests.

    We consider traces, generated using Intel's Pin Tool~\cite{PIN}, from the \texttt{GEMM}~\cite{GEMM}, \texttt{STREAM}~\cite{STREAM} benchmarks and Breadth First Search (\texttt{BFS}).\texttt{GEMM}, represents dense linear algebra operations, while \texttt{STREAM} consists of vector-based operations. Both demonstrate computational patterns that are characteristic of ML workloads. Additionally, we use traces from the SPEC CPU 2017~\cite{spec2017} suite, specifically high memory intensive applications, namely, \texttt{fotonik\_3d\_s},  \texttt{mcf\_s}, \texttt{lbm\_s}, and \texttt{roms\_s} stress the memory hierarchy due to their large data sets and frequent memory accesses. The compute intensive workloads include \texttt{xalancbmk\_s} and \texttt{gcc\_s}, which involve heavy computation for tasks such as XML transformations and code compilation. The \texttt{omnetpp\_s} requires intensive processing for network simulations while handling large amounts of simulation data, placing equal strain on the CPU and memory system.
    
    The SPEC CPU 2017 traces are generated using the ChampSim~\cite{gober2022championshipsimulatorarchitecturalsimulation} simulator, the traces are captured by monitoring last-level cache misses during simulations that execute at least ten billion instructions. The DRAMSys simulator integrated with DRAMPower~\cite{DRAMPower} provides performance metrics such as latency, average power consumption, total energy usage and, average and maximum bandwidth, etc. To gain deeper insights into memory behavior, we also extract additional metrics, including the number of bank group switches which occur when the memory controller switches between different bank groups within the DRAM, bank switches refers to switching between different banks within a bankgroup. Additionally, we track row buffer hits, which represent instances where the requested data is already in the row buffer, while row buffer misses occur when data is not in the buffer, requiring additional time to fetch from the corresponding row.
    
\subsection{Results}
        \begin{table*}[!t]
            \centering
            \renewcommand{\arraystretch}{1.5}
            \setlength{\tabcolsep}{6pt}
            \footnotesize
            \begin{tabular}{|l|c|c|c|c|c|c|c|}
                \hline
                \textbf{Workload} & 
                \textbf{Time Steps} & 
                \makecell{\textbf{Threshold} \\ \textbf{\(w\)}} & 
                \makecell{\textbf{Baseline} \\ \textbf{Reward}} & 
                \makecell{\textbf{ReLMXEL} \\ \textbf{Reward}} & 
                \makecell{\textbf{Average} \\ \textbf{Energy (\%)}} & 
                \makecell{\textbf{Average} \\ \textbf{Bandwidth (\%)}} & 
                \makecell{\textbf{Average} \\ \textbf{Latency (\%)}} \\
                \hline
                \textbf{STREAM} & 20170 & 16000 & 15555.06 & \textbf{17597.07} & 3.84 & 8.39 & 0.23 \\
                \hline
                \textbf{GEMM} & 19468 & 17000 & 6572.88 & \textbf{7121.46} & 3.83 & 4.95 & 0.01 \\
                \hline
                \textbf{BFS} & 17995 & 14000 & 9673.14 & \textbf{10842.41} & 7.66 & 7.22 & -0.03 \\
                \hline
                \textbf{fotonik\_3d} & 20770 & 17000 & 4870.89 & \textbf{9165.52} & 7.66 & 2.90 & 0.07 \\
                \hline
                \textbf{xalancbmk} & 16494 & 14000 & 3092.9 & \textbf{3320.38} & 7.68 & 107.03 & -0.02 \\
                \hline
                \textbf{gcc} & 17863 & 14000 & 9154.29 & \textbf{9556.25} & 7.66 & 1.70 & -0.24 \\
                \hline
                \textbf{roms} & 17563 & 14000 & 8017.8 & \textbf{13554.84} & 7.67 & 35.63 & 0.08 \\
                \hline
                \textbf{mcf} & 17894 & 14000 & 6013.5 & \textbf{6075.53} & 7.67 & 40.19 & -4.43 \\
                \hline
                \textbf{lbm} & 18473 & 15000 & 5496.77 & \textbf{14934.6} & 7.67 & 26.73 & 0.05 \\
                \hline
                \textbf{omnetpp} & 16682 & 14000 & 4743.99 & \textbf{6688.05} & 4.06 & 138.78 & -0.09\\
                \hline
                \end{tabular}
                \vspace*{0.25cm}
                \caption{Comparison of Baseline and ReLMXEL performance}
                \vspace*{-0.6cm}
                \label{tab:results_ReLMXEL}
        \end{table*}
        The experiments use a discount factor ($\gamma$) of 0.9 and a learning rate ($\alpha$) of 0.1. These values are chosen based on design space exploration across $\gamma \in \{0.9, 0.95, 0.99\}$ and $\alpha \in \{0.01, 0.1, 0.3, 0.5, 0.6, 0.7, 0.8\}$. While each workload has its own optimal ($\gamma$, $\alpha$) pair, the combination providing the highest reward across all workloads is used for all subsequent evaluations. We also introduce a \textit{Trace-split} parameter, that segments the trace file into fixed-size partitions. After each partition, the model makes decisions about the parameters and takes feedback from the SARSA using reward vector and Q-Tables, improving performance for the next timestep.
        
        Through experimentation, we set the trace split parameter to 30,000 and the exploration parameter $\epsilon_{\text{new}}$ to $0.001$, as values like $0.01$ hinder convergence due to excessive randomness, and $0.0001$ limit exploration, slowing recovery from suboptimal choices. The percentage improvements are computed relative to the baseline as follows: for energy and latency metrics, the improvement is calculated as
        \[
        \text{Improvement (\%)} = \frac{\text{Baseline} - \text{ReLMXEL}}{\text{Baseline}} \times 100,
        \] 
        so that a positive value indicates a reduction compared to the baseline. For the bandwidth metric, the improvement is calculated as 
        \[
        \text{Improvement (\%)} = \frac{\text{ReLMXEL} - \text{Baseline}}{\text{Baseline}} \times 100,
        \] 
        so that a positive value indicates an increase compared to the baseline.
        
        \begin{figure}[!h]
        \centering
        \begin{tikzpicture}
            \begin{axis}[
                    ybar,
                    bar width=5pt,
                    width=0.475\textwidth,
                    height=0.225\textwidth,
                    xlabel={Workloads},
                    ylabel={\small Avg Energy (pJ/$10^9$)},
                    symbolic x coords={STREAM,GEMM,BFS,fotonik\_3d,xalancbmk,gcc,roms,mcf,lbm,omnetpp},
                    xtick=data,
                    enlarge x limits=0.1,
                    ymin=0,
                    ymax=1800,
                    ymajorgrids=true,
                    legend style={at={(1,1)}, anchor=north east, fill=white, draw=black, draw=none, font=\footnotesize},
                    bar shift=0pt,
                    legend image code/.code={%
                        \draw[fill=#1, draw=none] (0cm,0cm) rectangle (0.1cm,0.1cm);
                    },
                    x tick label style={rotate=45, anchor=east},
                    xlabel style={yshift=-0.8cm},
                    ylabel style={xshift=-0.25cm,yshift=-0.1cm},
                    ]
                \addplot+[ybar, bar shift=-2pt, fill=black, draw=none] coordinates 
                    {(STREAM,951.9) (GEMM,959.58) (BFS,842.28) (fotonik\_3d,838.8) (xalancbmk,881.11) 
                     (gcc,839.76) (roms,807.67) (mcf,831.03) (lbm,803.1) (omnetpp,994.39)};
                \addplot+[ybar, bar shift=3pt, fill=gray, draw=none] coordinates 
                    {(STREAM,915.38) (GEMM,922.81) (BFS,777.72) (fotonik\_3d,774.52) (xalancbmk,813.39) 
                     (gcc,775.41) (roms,745.74) (mcf,767.29) (lbm,741.48) (omnetpp,953.96)};
                \legend{Baseline,ReLMXEL}
            \end{axis}
            \end{tikzpicture}
            \vspace*{-0.3cm}
            \caption{Average energy consumption}
            \vspace*{-0.3cm}
            \label{fig:avg_energy}
        \end{figure}
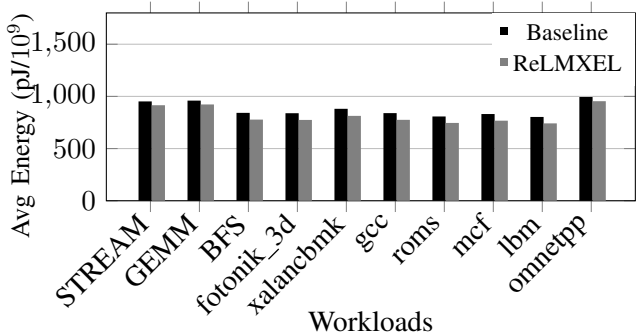

        The $\%$ improvement of average Energy, Bandwidth and Latency columns in Table~\ref{tab:results_ReLMXEL} show that ReLMXEL consistently outperforms the baseline across all workloads. ReLMXEL achieves high bandwidth utilization and reduced latency, while also exhibiting slightly better energy efficiency than the baseline in memory-bound workloads, such as \texttt{STREAM} and \texttt{GEMM}. It also performs well in bandwidth utilization and energy efficiency for irregular and graph-based workloads, including \texttt{BFS}, \texttt{fotonik\_3d}, and \texttt{roms}, as well as on compute-intensive workloads, such as \texttt{xalancbmk}, \texttt{gcc}, and \texttt{lbm}, reflecting optimized computation scheduling. Workloads with high memory traffic or communication demands, including \texttt{mcf} and \texttt{omnetpp}, achieve improvement in energy consumption and bandwidth utilization; however a slight increase in latency, indicates a trade-off between energy efficiency and data transfer overhead.

        \begin{figure}[!h]
        \centering
        \begin{tikzpicture}
            \begin{axis}[
                    ybar,
                    bar width=5pt,
                    width=0.475\textwidth,
                    height=0.225\textwidth,
                    xlabel={Workloads},
                    ylabel={\small Avg Bandwidth (Gb/s)},
                    symbolic x coords={STREAM,GEMM,BFS,fotonik\_3d,xalancbmk,gcc,roms,mcf,lbm,omnetpp},
                    xtick=data,
                    enlarge x limits=0.1,
                    ymin=0,
                    ymax=150,
                    ymajorgrids=true,
                    legend style={at={(1,1)}, anchor=north east, fill=white, draw=black, draw=none, font=\footnotesize},
                    bar shift=0pt,
                    legend image code/.code={%
                        \draw[fill=#1, draw=none] (0cm,0cm) rectangle (0.1cm,0.1cm);
                    },
                    x tick label style={rotate=45, anchor=east},
                    xlabel style={yshift=-0.8cm},
                    ylabel style={xshift=-0.2cm,yshift=-0.3cm},
                    ]
                \addplot+[ybar, bar shift=-2pt, fill=black, draw=none] coordinates 
                    {(STREAM,81.24) (GEMM,73.47) (BFS,70.81) (fotonik\_3d,69.68) (xalancbmk,44.52) 
                     (gcc,42.59) (roms,52.63) (mcf,59.31) (lbm,23.94) (omnetpp,26.82)};
                \addplot+[ybar, bar shift=3pt, fill=gray, draw=none] coordinates 
                    {(STREAM,88.06) (GEMM,77.11) (BFS,75.92) (fotonik\_3d,71.7) (xalancbmk,92.17) 
                     (gcc,43.31) (roms,71.38) (mcf,83.15) (lbm,30.34) (omnetpp,64.04)};
                \legend{Baseline,ReLMXEL}
            \end{axis}
            \end{tikzpicture}
            \vspace*{-0.3cm}
            \caption{Average bandwidth utilization}
            \vspace*{-0.5cm}
            \label{fig:avg_bandwidth}
        \end{figure}
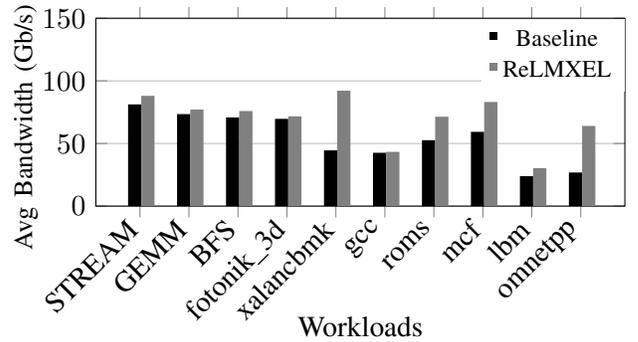
        
        \begin{figure}[!h]
        \centering
        \begin{tikzpicture}
            \begin{axis}[
                    ybar,
                    bar width=5pt,
                    width=0.475\textwidth,
                    height=0.225\textwidth,
                    xlabel={Workloads},
                    ylabel={\small Avg Latency (ps/$10^6$)},
                    symbolic x coords={STREAM,GEMM,BFS,fotonik\_3d,xalancbmk,gcc,roms,mcf,lbm,omnetpp},
                    xtick=data,
                    enlarge x limits=0.1,
                    ymin=0,
                    ymax=5000000,
                    ymajorgrids=true,
                    legend style={at={(1,1)}, anchor=north east, fill=white, draw=black, draw=none, font=\footnotesize},
                    bar shift=0pt,
                    legend image code/.code={%
                        \draw[fill=#1, draw=none] (0cm,0cm) rectangle (0.1cm,0.1cm);
                    },
                    x tick label style={rotate=45, anchor=east},
                    xlabel style={yshift=-0.8cm},
                    ylabel style={xshift=0cm,yshift=-0.4cm},
                    ]
                \addplot+[ybar, bar shift=-2pt, fill=black, draw=none] coordinates 
                    {(STREAM,2693460.34) (GEMM,2715493.46) (BFS,2384200.57) (fotonik\_3d,2373252.05) (xalancbmk,2490541.25) (gcc,2368234.94) (roms,2284908.97) (mcf,2248064.97) (lbm,2268539.66) (omnetpp,2809860.91)};
                \addplot+[ybar, bar shift=3pt, fill=gray, draw=none] coordinates 
                    {(STREAM,2687368.2) (GEMM,2715343.55) (BFS,2384919.40) (fotonik\_3d,2371580.82) (xalancbmk,2491011.58) (gcc,2373888.01) (roms,2283034.71) (mcf,2347576.55) (lbm,2267302.82) (omnetpp,2812492.07)};
                \legend{Baseline,ReLMXEL}
            \end{axis}
            \end{tikzpicture}
            \vspace*{-0.3cm}
            \caption{Average latency}
            \vspace*{-0.1cm}
            \label{fig:avg_latency}
        \end{figure}
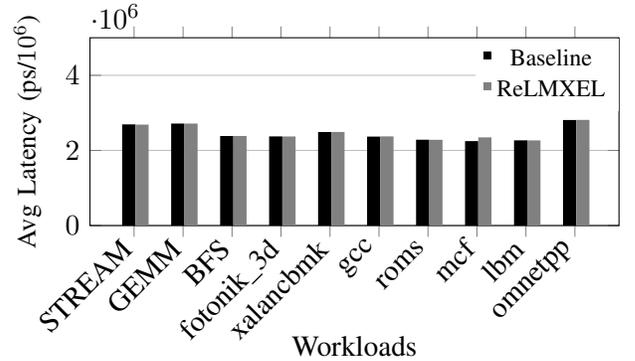

        Figures~\ref{fig:avg_energy}, ~\ref{fig:avg_bandwidth}, and ~\ref{fig:avg_latency} illustrate how ReLMXEL’s dynamic tuning approach incrementally optimizes memory controller parameters through a step-by-step, feedback driven process that adapts to real time workload characteristics. Leveraging a multi-agent reinforcement learning framework with explainability, it balances competing objectives to optimize overall system performance. As a result, significant reductions in energy consumption and bandwidth gains are achieved across diverse workloads, particularly in memory-bound and irregular patterns, without causing substantial latency degradation. This minimal impact on latency demonstrates that ReLMXEL successfully navigates the tradeoffs inherent in system optimization, proving the effectiveness of its adaptive, feedback driven parameter tuning in delivering balanced and robust performance improvements.

    \section{Conclusion and Future Directions}
        The proposed ReLMXEL based memory controller achieved enhanced efficiency and transparency. The RL framework proposed optimizes memory controller parameters while decomposing rewards to model energy, bandwidth, and latency trade-offs. Experimental results showed significant performance improvements across diverse workloads, confirming the framework's ability to balance competing system objectives. This integration of adaptive learning with interpretable decision-making marks a key advancement in memory systems, paving the way for future research into self-optimizing, high-performance architectures with explainability.
    
        As RL optimizes memory controller parameters, enabling adaptive responses to dynamic workloads, RL based optimizations can be extended to heterogeneous memory architectures, such as hybrid nonvolatile memory systems, to assess its robustness in real-world scenarios. Integrating RL with hardware in the loop setups allows real time interaction with actual hardware, bridging the gap between simulations and real-world applications. Additionally, RL can help in efficient detection and mitigation of DRAM security threats like row hammer attacks, by identifying malicious memory access patterns and adjusting memory access strategies to prevent data corruption or security breaches.

\bibliographystyle{IEEEtran}
\bibliography{references}

@book{rlanintroduction,
author = {Sutton, Richard S. and Barto, Andrew G.},
title = {Reinforcement Learning: An Introduction},
year = {2018},
isbn = {0262039249},
publisher = {A Bradford Book},
address = {Cambridge, MA, USA}
}

@article{DRAMSys,
  author = {Steiner, Lukas and Jung, Matthias and Prado, Fernando S. and others},
  title = {DRAMSys4.0: An Open-Source Simulation Framework for In-depth DRAM Analyses},
  journal = {International Journal of Parallel Programming},
  volume = {50},
  pages = {217--242},
  year = {2022},
  doi = {10.1007/s10766-022-00727-4}
}

@article{survey,
   title={A Survey of Machine Learning for Computer Architecture and Systems},
   volume={55},
   ISSN={1557-7341},
   url={http://dx.doi.org/10.1145/3494523},
   DOI={10.1145/3494523},
   number={3},
   journal={ACM Computing Surveys},
   publisher={Association for Computing Machinery (ACM)},
   author={Wu, Nan and Xie, Yuan},
   year={2022},
   month=feb, pages={1-39} }

@article{Q-Table,
  author = {Watkins, Christopher J. C. H. and Dayan, Peter},
  day = 01,
  doi = {10.1007/BF00992698},
  issn = {1573-0565},
  journal = {Machine Learning},
  month = may,
  number = 3,
  pages = {279--292},
  title = {Q-learning},
  url = {https://doi.org/10.1007/BF00992698},
  volume = 8,
  year = 1992
}

@inproceedings{juozapaitis2019rewarddecomposition,
  author = {Juozapaitis, Z and Koul, A and Fern, A and Erwig, M and Doshi-Velez, F},
  title = {Explainable Reinforcement Learning via Reward Decomposition},
  booktitle = {Proceedings of the International Joint Conference on Artificial Intelligence (IJCAI) Workshop on Explainable Artificial Intelligence},
  year = {2019}
}

@inproceedings{PIN,
author = {Reddi, Vijay Janapa and Settle, Alex and Connors, Daniel A. and Cohn, Robert S.},
title = {PIN: a binary instrumentation tool for computer architecture research and education},
year = {2004},
isbn = {9781450347334},
publisher = {Association for Computing Machinery},
address = {New York, NY, USA},
url = {https://doi.org/10.1145/1275571.1275600},
doi = {10.1145/1275571.1275600},
booktitle = {Proceedings of the 2004 Workshop on Computer Architecture Education: Held in Conjunction with the 31st International Symposium on Computer Architecture},
pages = {22-es},
location = {Munich, Germany},
series = {WCAE '04}
}

@book{rummerysarsa,
  title={On-line Q-learning using connectionist systems},
  author={Rummery, Gavin A and Niranjan, Mahesan},
  volume={37},
  year={1994},
  publisher={University of Cambridge, Department of Engineering Cambridge, UK}
}

@article{cmac,
  author = {James Albus},
  title = {New Approach to Manipulator Control: The Cerebellar Model Articulation Controller (CMAC)1},
  year = {1975},
  month = {1975-09-30},
  publisher = {Transactions of the ASME Journal of Dynamic Systems},
  url = {https://tsapps.nist.gov/publication/get_pdf.cfm?pub_id=820151},
  language = {en},
}

@inproceedings{Selfoptimising,
  author={Ipek, Engin and Mutlu, Onur and Martínez, José F. and Caruana, Rich},
  booktitle={2008 International Symposium on Computer Architecture}, 
  title={Self-Optimizing Memory Controllers: A Reinforcement Learning Approach}, 
  year={2008},
  volume={},
  number={},
  pages={39-50},
  doi={10.1109/ISCA.2008.21}}

@INPROCEEDINGS{FRFCFS,
  author={Rixner, S.},
  booktitle={37th International Symposium on Microarchitecture (MICRO-37'04)}, 
  title={Memory Controller Optimizations for Web Servers}, 
  year={2004},
  volume={},
  number={},
  pages={355-366},
  doi={10.1109/MICRO.2004.22}}

@inproceedings{Bera_2021,
  title     = {Pythia: A Customizable Hardware Prefetching Framework Using Online Reinforcement Learning},
  author    = {Rahul Bera and Konstantinos Kanellopoulos and Anant Nori and Taha Shahroodi and Sreenivas Subramoney and Onur Mutlu},
  booktitle = {MICRO-54: 54th Annual IEEE/ACM International Symposium on Microarchitecture},
  year      = {2021},
  month     = {oct},
  pages     = {1121--1137},
  publisher = {ACM},
  doi       = {10.1145/3466752.3480114},
  url       = {http://dx.doi.org/10.1145/3466752.3480114}
}

@article{STREAM,
  author    = {John D. McCalpin},
  title     = {Memory Bandwidth and Machine Balance in Current High Performance Computers},
  journal   = {IEEE Computer Society Technical Committee on Computer Architecture (TCCA) Newsletter},
  year      = {1995},
  month     = {dec},
  pages     = {19--25},
  note      = {\url{http://tab.computer.org/tcca/NEWS/DEC95/dec95_mccalpin.ps}}
}

@ARTICLE{gober2022championshipsimulatorarchitecturalsimulation,
       author = {{Gober}, Nathan and {Chacon}, Gino and {Wang}, Lei and {Gratz}, Paul V. and {Jimenez}, Daniel A. and {Teran}, Elvira and {Pugsley}, Seth and {Kim}, Jinchun},
        title = "{The Championship Simulator: Architectural Simulation for Education and Competition}",
      journal = {arXiv e-prints},
         year = 2022,
        month = oct,
          eid = {arXiv:2210.14324},
        pages = {arXiv:2210.14324},
          doi = {10.48550/arXiv.2210.14324},
archivePrefix = {arXiv},
       eprint = {2210.14324},
 primaryClass = {cs.AR},
       adsurl = {https://ui.adsabs.harvard.edu/abs/2022arXiv221014324G}
}

@misc{GEMM,
  title     = {GEMMbench: a framework for reproducible and collaborative benchmarking of matrix multiplication},
  author    = {Anton Lokhmotov},
  year      = {2015},
  eprint    = {1511.03742},
  archivePrefix = {arXiv},
  primaryClass  = {cs.MS},
  url       = {https://arxiv.org/abs/1511.03742}
}

@misc{intel2024openpage,
  author       = {{Intel Corporation}},
  title        = {Performance Differences for Open-Page / Close-Page Policy},
  year         = {2024},
  month        = jul,
  url          = {https://www.intel.com/content/www/us/en/content-details/826015/performance-differences-for-open-page-close-page-policy.html},
}

@misc{spec2017,
  title     = {SPEC CPU 2017 Benchmark Suite},
  author    = {{Standard Performance Evaluation Corporation}},
  year      = {2017},
  url       = {https://www.spec.org/cpu2017/}
}

@misc{DRAMPower,
  author       = {Karthik Chandrasekar and Christian Weis and Yonghui Li and Sven Goossens and Matthias Jung and Omar Naji and Benny Akesson and Norbert Wehn and Kees Goossens},
  title        = {{DRAMPower: Open-source DRAM Power \& Energy Estimation Tool}},
  howpublished = {\url{http://www.drampower.info}},
  note         = {Accessed: April 2025},
  year         = {2014}
}

@online{jedecDDR4,
  author =       "",
  year =         "2021",
  title =        "JEDEC DDR4 SDRAM Standard Document",
  url =          "https://www.jedec.org/standards-documents/docs/jesd79-4a",
  month =        July,
  lastaccessed = "July, 2021",
}

\end{document}